\documentclass[final]{raa}

\usepackage{natbib}
\usepackage{hyperref}
\usepackage{color}
\usepackage{natbib}
\usepackage{graphicx} 

\begin{document}

\title{Transit Timing Variations and linear ephemerides of confirmed Kepler transiting exoplanets}

\volnopage{Vol. \textbf{19} (2019) No. \textbf{3}, 41}      
\setcounter{page}{1}          

\author{Pavol Gajdo\v{s}
\inst{1}
\and Martin Va\v{n}ko
\inst{2}
\and \v{S}tefan Parimucha
\inst{1}
}

\institute{Institute of Physics, Faculty of Science, Pavol Jozef \v{S}af\'{a}rik University, Ko\v{s}ice, Slovakia; {\it pavol.gajdos@student.upjs.sk}\\
\and
Astronomical Institute, Slovak Academy of Sciences, 059 60 Tatransk\'a Lomnica, Slovakia}

\date{Received~~2018 June 28; accepted~~2018~~September 28}

\titlerunning{TTVs \& linear ephemerides of exoplanets}
\authorrunning{Gajdo\v{s} et al.}

\abstract{We determined new linear ephemerides of transiting exoplanets using long-cadence de-trended data from quarters Q1 to Q17 of Kepler mission.
We analysed TTV diagrams of 2098 extrasolar planets. The TTVs of 121 objects were excluded (because of insufficient data-points, influence of
stellar activity, etc). Finally, new linear ephemerides of 1977 exoplanets from Kepler archive are presented. The significant linear trend was observed on TTV diagrams of approximately 35\% of studied exoplanets. Knowing correct linear ephemeris is principal for successful follow-up observations of transits. Residual TTV diagrams of 64 analysed exoplanets shows periodic variation, 43 of these TTV planets were not reported yet.
}

\maketitle

\section{Introduction}
The Kepler satellite, launched in 2009, provided during its primary mission high-precision, high-cadence and continuous photometric data \citep{borucki2010}. After losing two reaction wheels in 2013, so-called K2 mission started and still continue \citep{Howell2014}. 

During the primary mission, Kepler discovered 2327 extrasolar planets (up to May 31$^{st}$, 2018). Almost the half of them (1125) are located in 447 multi-planet systems. The final catalogue (DR25) of Kepler planet candidate was released in 2017 \citep{Thompson2018}. It consists of more than four thousand of planet candidates.

In many of known exoplanets, the variations in times of transits were already observed. \cite{Holczer2016} detected 260 planet candidates with significant long-term variations. These variations could be caused by gravitational interaction with another bodies in the system. For example, \cite{Steffen2012} determined the masses of planets in the systems Kepler-25, Kepler-26, Kepler-27 and Kepler-28 using transit timing. Similar method is also used to confirm the planets in multi-planetary systems (e.g. \cite{Fabrycky2012}). Many of planet pairs are captured into mean motion resonances (MMRs) \citep{Wang2017}. The MMRs 3:2 and 2:1 are the most common\citep{Wang2014}.

\begin{table*}
\caption{The new linear ephemerides of Kepler exoplanets: Kepler, KOI (Kepler Object of Interest) and KIC (Kepler Input Catalog ) - name of exoplanet, $P$ - orbital period, $T_0$ - an initial time of transit (uncertainties are given in parenthesis), $\chi^2$ - sum of squares of the best fit, $\chi^2/n$ - reduced sum of squares, $n$ - number of data points in TTV diagram, Note - changes of residual TTV after new ephemeris is removed, 1 - no significant changes, 2 - periodic or quasi-periodic variations, 3 - chaotic variations (see text for a detailed description). The full table is available as a supplementary material to this manuscript.}
\label{tab:ephm}
\begin{center}
{\scriptsize
\begin{tabular}{ccc|cc|ccc|c}
	\hline\hline
	    \multicolumn{3}{c|}{Names}     & \multicolumn{2}{c|}{New ephemeris} & \multicolumn{3}{c|}{Fit statistics} & Note  \\
	   Kepler     &   KOI   &   KIC    &   $P$ (d)    &     $T_0$ (BJD)     & $\chi^2$ & $\chi^2/n$ &     $n$     &  \\ \hline
	 Kepler-4 b   &  7.01   & 11853905 & 3.213663(1)  &  2454956.61167(24)  &  523.9   &    1.6     &     330     &   1   \\
	 Kepler-5 b   &  18.01  & 8191672  & 3.5484659(1) &  2454955.90135(3)   &  6020.5  &    16.0    &     379     &   1   \\
	 Kepler-6 b   &  17.01  & 10874614 & 3.2346994(8) &  2454954.48659(2)   &  5598.0  &    16.8    &     335     &   1   \\
	    \dots     &  \dots  &  \dots   &    \dots     &        \dots        &  \dots   &   \dots    &    \dots    & \dots \\
	Kepler-1646 b & 6863.01 & 7350067  & 4.485563(2)  &  2454968.43834(30)  & 9126.425 &   31.579   &     291     &   1   \\ \hline
\end{tabular}}
\end{center}	
\end{table*}

\begin{figure*}[t]
\begin{center}
\includegraphics[width=0.9\textwidth]{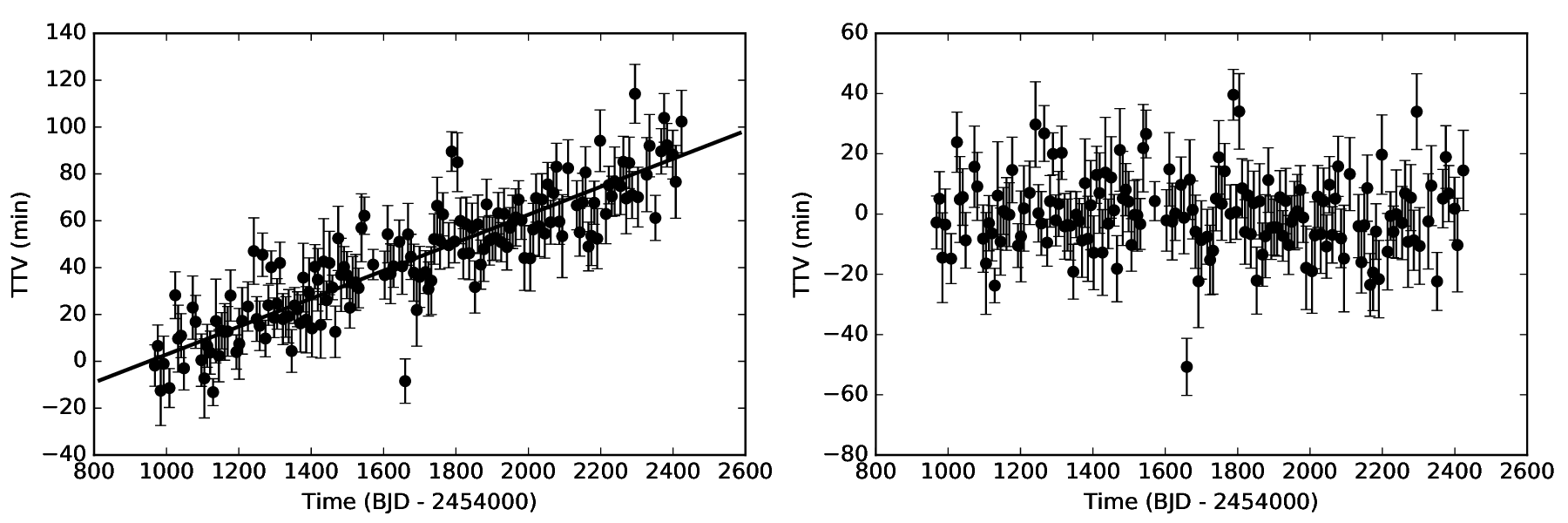}
\caption{TTV of Kepler-114~c determined from original ephemeris from NASA Exoplanet Archive with new linear fit (solid line), $right$ --  residual TTV of Kepler-114~c (note 1 in Table~\ref{tab:ephm}).}
\label{fig:oc-lin}
\end{center}
\end{figure*}

\section{ Linear ephemeris determination}
\label{time}
In our analysis we used long-cadence (sampled every 29.4 minutes) de-trended data (PDCSAP\_FLUX) from quarters Q1 to Q17 of Kepler mission, obtained from Mikulski Archive for Space Telescopes (MAST)\footnote{doi:10.17909/T9059R}. For the analysis of data, we used the same pipeline as in our study of transit-timing variations (TTVs) in the system Kepler-410 \citep{kep410}. We used same approach for all studied planets and got homogeneous set of times of transits by one method. It can be summarized in the following steps:
\begin{enumerate}
\item Parts of the light curve (LC) of the individual system were extracted around detected transits (using ephemeris given by NASA Exoplanet Archive\footnote{We used Confirmed Planets table available at\\ https://exoplanetarchive.ipac.caltech.edu/cgi-bin/TblView/nph-tblView?app=ExoTbls\&config=planets. Downloaded at April 2018} \citep{akeson2013}), with an interval two times bigger than the transit duration.
\item Additional residual trends caused by the stellar activity and/or instrumental long-term photometric variation were removed by the fitting out-of-transit part of LC by the second-order polynomial function. 
\item All individual parts of the LC with transits were stacked together. This can be done, because one expects that the physical parameters of the host star and the exoplanet did not change during the observational period of about 3.5 years and we want to cancel-out the effect of stellar activity.
\item Stacked LC was fitted by our software implementation of \citet{MA} model where we used a quadratic model of limb darkening with values of coefficients from \citet{Sing}. Our package use  Markov Chain Monte-Carlo (MCMC) simulation to obtain statistically significant value of parameters and their errors. 
\item Obtained template was used to fit all individual transits, where only the time of transit was updated. 
\item Determined times of transits  were used for a creation of transit timing variations (TTV) diagram of the object. It was subsequently fitted by the linear function to obtain new values of linear ephemeris parameters, initial time of transit $T_0$ and orbital period $P$. To achieve statistically significant estimation of parameter's uncertainties, we used MCMC simulation. To estimate the quality of the statistical model, we have calculated sum of squares $\chi^2$ and reduced sum of squares $\chi^2/n$ 
where $n$ is the number of data points in TTV diagram.
\item Finally, new linear trend determined by new ephemeris was removed and residual TTV was visually inspected for another changes.
\end{enumerate}

\begin{table*}
\caption{Comparison of periods and amplitudes of TTV changes found in this paper with values given by \cite{Holczer2016}: Kepler, KOI  and KIC - name of exoplanet, $P$ - period of variation, $A$ - amplitude (uncertainties are given in parenthesis).}
\label{tab:ttv-old}
\begin{center}
{\scriptsize
\begin{tabular}{ccc|cc|cc}
	\hline\hline
	    \multicolumn{3}{c|}{Names}     & \multicolumn{2}{c|}{\cite{Holczer2016}} & \multicolumn{2}{c}{This paper} \\
	   Kepler     &   KOI   &   KIC    &  $P$ (d)  &           $A$ (m)           &  $P$ (d)  &      $A$ (m)       \\ \hline
	 Kepler-25 b  & 244.02  & 4349452  &  326(3)   &           3.8(3)            &  334(2)   &       4.3(3)       \\
	 Kepler-51 b  & 620.01  & 11773022 &  790(12)  &           7.9(4)            &  749(17)  &       5.5(5)       \\
	 Kepler-81 c  & 877.02  & 7287995  &  536(12)  &            9(2)             &  516(9)   &       10(1)        \\
	Kepler-111 c  & 139.01  & 8559644  & 2213(79)  &           211(22)           & 1050(20)  &       38(2)        \\
	Kepler-139 c  & 316.02  & 8008067  & 1008(84)  &            29(8)            &  913(30)  &       35(3)        \\
	Kepler-209 c  & 672.02  & 7115785  & 1061(76)  &            9(2)             & 1141(63)  &       10(1)        \\
	Kepler-221 e  & 720.03  & 9963524  &    ---    &           $\sim$9           & 1697(115) &       11(1)        \\
	Kepler-278 c  & 1221.02 & 3640905  &  829(41)  &           144(32)           &  960(69)  &      112(16)       \\
	Kepler-312 c  & 1628.01 & 6975129  &  471(17)  &            6(2)             &  478(14)  &        7(1)        \\
	Kepler-359 c  & 2092.01 & 6696580  & 1270(130) &            23(6)            & 1246(125) &       20(6)        \\
	Kepler-540 b  & 374.01  & 8686097  & 1388(84)  &            38(6)            & 1457(54)  &       37(4)        \\
	Kepler-561 b  & 464.01  & 8890783  &  482(11)  &           3.9(6)            &  480(12)  &       4.1(7)       \\
	Kepler-591 b  & 536.01  & 10965008 &  454(37)  &            6(4)             &  579(35)  &        9(2)        \\
	Kepler-765 b  & 1086.01 & 10122255 & 1630(150) &            20(4)            & 1445(133) &       37(3)        \\
	Kepler-827 b  & 1355.01 & 7211141  &  124(1)   &            6(2)             &  398(36)  &        6(1)        \\
	Kepler-1040 b & 1989.01 & 10779233 &    ---    &          $\sim$15           &  901(30)  &       17(4)        \\
	Kepler-1624 b & 4928.01 & 1873513  &  110(1)   &           2.1(7)            &  118(3)   &       1.9(2)       \\ \hline
\end{tabular}}
\end{center}
\end{table*}

\begin{table*}
\caption{Planets with periodic variations in TTV diagram not reported in \cite{Holczer2016}.}
\label{tab:ttv}
\begin{center}
{\scriptsize
\begin{tabular}{ccc|cc}
	\hline\hline
	    \multicolumn{3}{c|}{Names}     & \multicolumn{2}{c}{Variation} \\
	   Kepler     &   KOI   &   KIC    &  $P$ (d)  &      $A$ (m)      \\ \hline
	 Kepler-39 b  & 423.01  & 9478990  & 1393(81)  &      1.9(3)       \\
	 Kepler-52 d  & 775.03  & 11754553 & 1493(300) &       9(4)        \\
	 Kepler-89 c  &  94.02  & 6462863  &  395(10)  &       10(2)       \\
	Kepler-110 b  & 124.01  & 11086270 &  767(23)  &       12(4)       \\
	Kepler-110 c  & 124.02  & 11086270 & 1245(13)  &       6(2)        \\
	Kepler-122 e  & 232.04  & 4833421  & 1135(59)  &       42(6)       \\
	Kepler-142 d  & 343.03  & 10982872 &  1400(5)  &       57(9)       \\
	Kepler-166 c  & 481.03  & 11192998 & 1419(81)  &       20(5)       \\
	Kepler-196 c  & 612.02  & 6587002  & 1431(197) &       7(1)        \\
	Kepler-201 c  & 655.02  & 5966154  &  828(88)  &       11(4)       \\
	Kepler-222 d  & 723.02  & 10002866 &  770(23)  &       5(1)        \\
	Kepler-222 c  & 723.03  & 10002866 &  606(11)  &      3.2(5)       \\
	Kepler-227 c  & 752.02  & 10797460 & 1041(12)  &       14(8)       \\
	Kepler-230 c  & 759.02  & 11018648 &  776(60)  &       19(6)       \\
	Kepler-233 c  & 790.02  & 12470844 & 1057(127) &       12(5)       \\
	Kepler-267 d  & 1078.03 & 10166274 &  818(24)  &       8(2)        \\
	Kepler-283 c  & 1298.02 & 10604335 & 1212(36)  &      68(11)       \\
	Kepler-299 e  & 1432.04 & 11014932 & 1427(102) &      74(46)       \\
	Kepler-300 c  & 1435.01 & 11037335 & 1365(104) &      58(10)       \\
	Kepler-310 c  & 1598.01 & 10004738 & 1004(10)  &       5(2)        \\
	Kepler-310 b  & 1598.03 & 10004738 & 1134(49)  &       15(3)       \\
	Kepler-358 c  & 2080.01 & 10864531 &  810(3)   &       19(6)       \\
	Kepler-362 c  & 2147.01 & 10404582 &  851(85)  &       14(6)       \\
	Kepler-364 c  & 2153.01 & 10253547 &  696(66)  &       23(4)       \\
	Kepler-509 b  & 276.01  & 11133306 & 1506(138) &       4(1)        \\
	Kepler-549 c  & 427.03  & 10189546 &  978(43)  &       42(8)       \\
	Kepler-672 b  & 773.01  & 11507101 &  707(32)  &       14(3)       \\
	Kepler-795 b  & 1218.01 & 3442055  &  731(27)  &       10(4)       \\
	Kepler-797 b  & 1238.01 & 6383821  &  895(60)  &      37(11)       \\
	Kepler-807 b  & 1288.01 & 10790387 & 1327(97)  &      1.8(5)       \\
	Kepler-852 b  & 1444.01 & 11043167 & 1123(76)  &       15(3)       \\
	Kepler-966 b  & 1828.01 & 11875734 & 1162(120) &       7(2)        \\
	Kepler-1036 b & 1980.01 & 11769890 & 1382(84)  &       16(2)       \\
	Kepler-1097 b & 2102.01 & 7008211  &  729(29)  &       9(1)        \\
	Kepler-1126 b & 2162.01 & 9205938  & 1327(133) &       23(6)       \\
	Kepler-1129 c & 2167.03 & 6041734  & 1236(117) &       19(4)       \\
	Kepler-1184 b & 2309.01 & 10010440 & 1108(51)  &       24(5)       \\
	Kepler-1185 b & 2311.03 & 4247991  &  820(89)  &       9(3)        \\
	Kepler-1388 b & 2926.01 & 10122538 &  370(19)  &       18(7)       \\
	Kepler-1389 b & 2931.01 & 8611257  & 1312(101) &       43(8)       \\
	Kepler-1453 b & 3280.01 & 10653179 &  1078(2)  &      22(10)       \\
	Kepler-1524 b & 3878.01 & 4472818  &  974(77)  &       4(1)        \\
	Kepler-1527 b & 3901.01 & 9480535  & 1024(25)  &       44(5)       \\
	Kepler-1530 c & 3925.03 & 10788461 &  86.4(6)  &       13(3)       \\
	Kepler-1552 b & 4103.01 & 3747817  & 998(125)  &       28(8)       \\
	Kepler-1593 b & 4356.01 & 8459663  & 1465(134) &       28(9)       \\
	Kepler-1638 b & 5856.01 & 11037818 &  840(51)  &      197(32)      \\ \hline
\end{tabular}}
\end{center}
\end{table*}

\begin{figure*}[t]
\begin{center}
\includegraphics[width=0.9\textwidth]{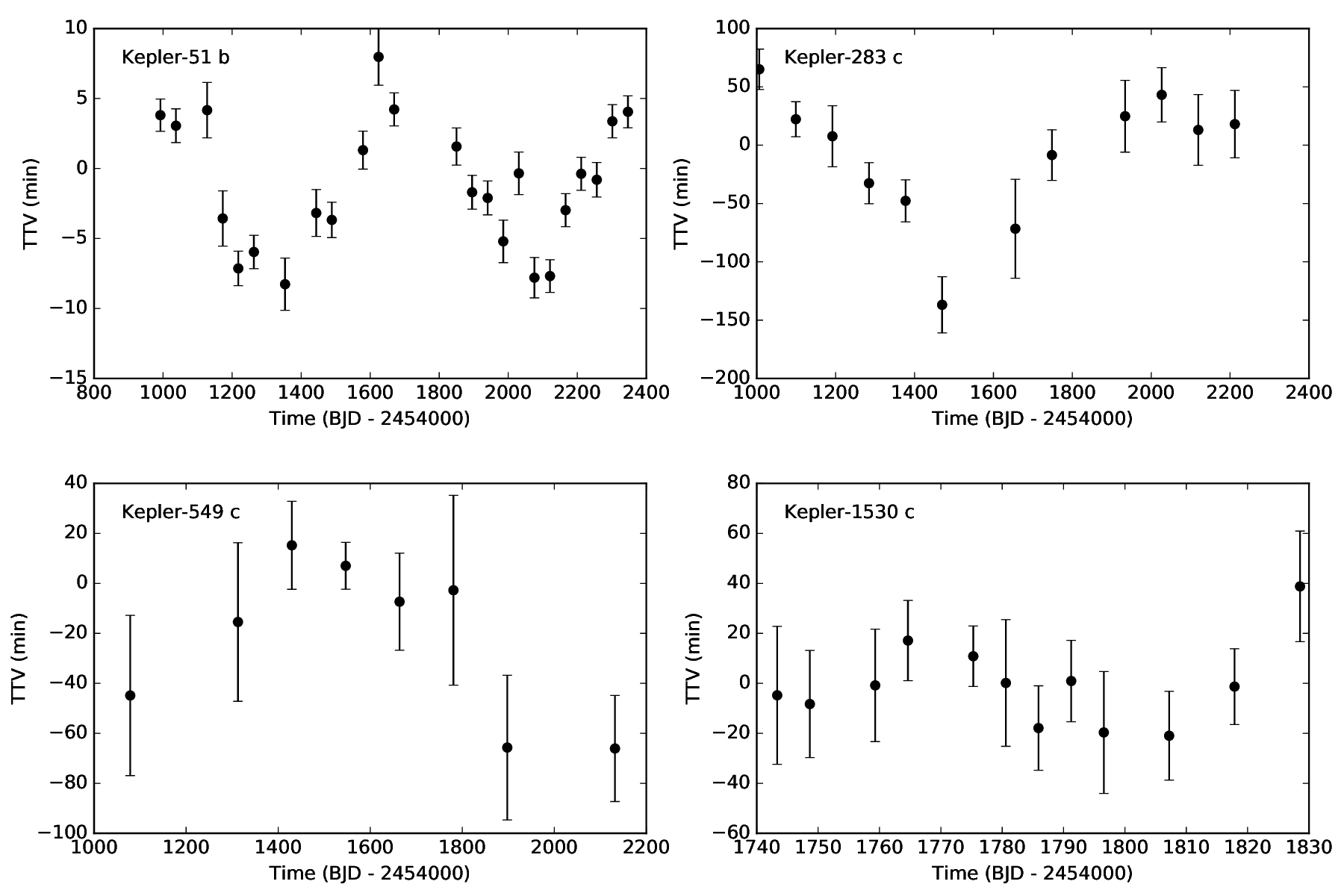}
\caption{Examples of planets with periodic variations on the residual TTV diagram (note 2 in Table~\ref{tab:ephm}). Periods and amplitudes of these changes are listed in Tables~\ref{tab:ttv-old} and \ref{tab:ttv}.}
\label{fig:oc-ttv}
\end{center}
\end{figure*}

\begin{figure*}[t]
\begin{center}
\includegraphics[width=0.9\textwidth]{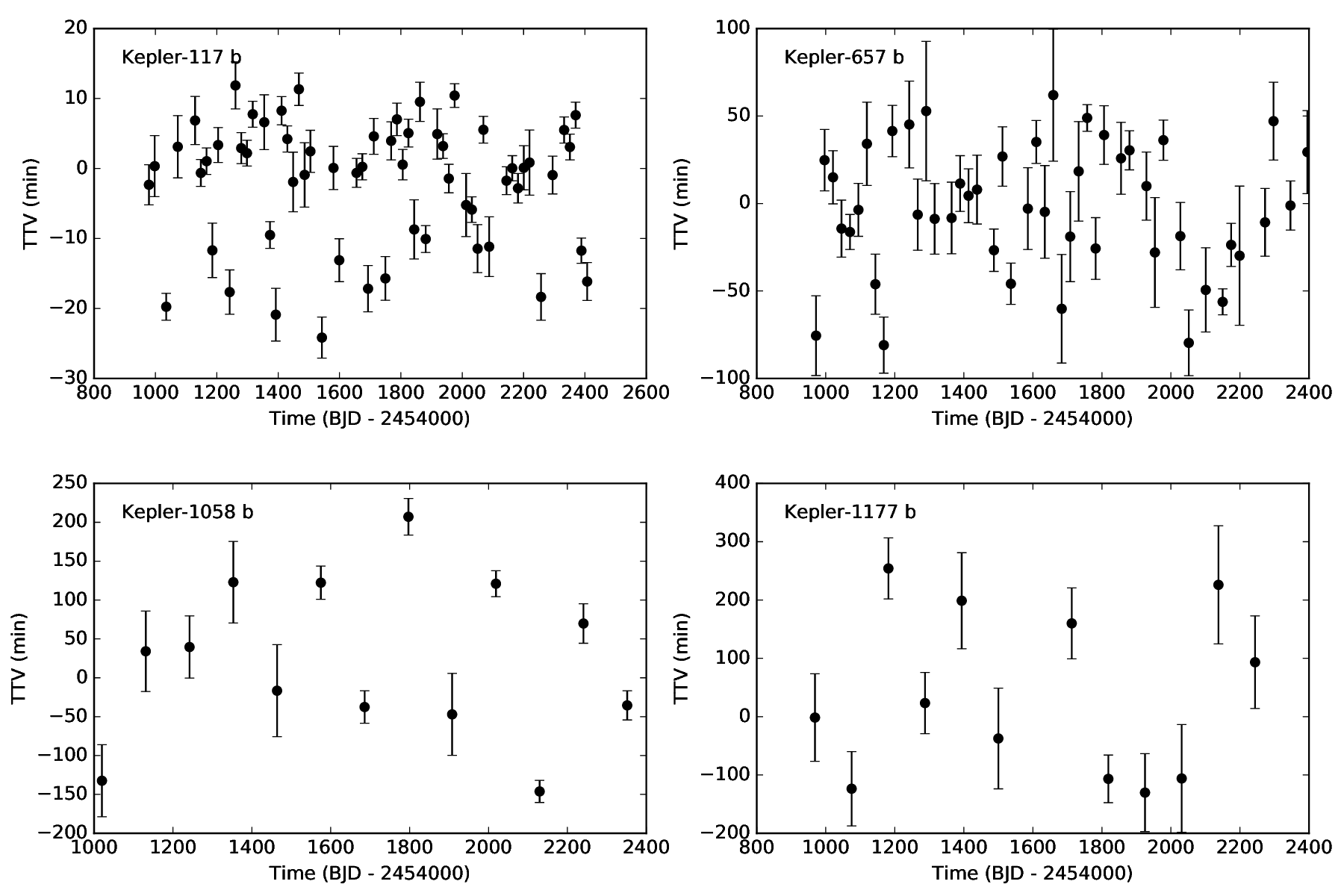}
\caption{Examples of exoplanets with chaotic (abnormal) TTVs (note 3 in Table~\ref{tab:ephm}).}
\label{fig:oc-chaotic}
\end{center}
\end{figure*}

\section{Discussion and conclusion}
\label{oc}
We started our analysis of TTV diagrams with 2098 extrasolar planets from Kepler database. We excluded 121 objects with TTV diagrams consisting less than 4 points or with diagrams strongly affected by stellar activity. New linear ephemeris were determined for 1977 exoplanets. They are all presented in Table~\ref{tab:ephm}.

Our analysis revealed that in many cases ($\sim 35\%$), linear trend could be observed on the TTV diagrams which were calculated according to original ephemerides given by NASA Exoplanet Archive. Cumulative shift in the minima times of studied exoplanets can reach up to $\sim 130$ minutes (e.g. Kepler-4~b) over 3.5 years of Kepler observations. The example of such a significant trend detected in the system Kepler-114~c is shown on Fig.~\ref{fig:oc-lin} (left). The period of Kepler-114~c given by the NASA Exoplanet Archive is 8.041 days. But the reference \citep{Xie2014} for it is quite old and had used only data up to Q16 quarter. We used also Q17 data in this paper. Also in many other cases, the ephemeris given by the Archive is not the latest one. The incorrect value of period could cause that the transit will be really observed few hours earlier or latter than it will be calculated, after few years. And the observer with outdated ephemeris will not see any transit at all. After removing linear trend determined by the new ephemeris, we can obtain residual TTV with no other significant changes (right) (note 1 in Table~\ref{tab:ephm}).

Residual TTV of 64 planets shows periodic or quasi-periodic variations (note 2 in Table~\ref{tab:ephm}). The examples of such systems with more or less significant changes are depicted in Fig.~\ref{fig:oc-ttv}. TTV of 17 planets from this group were already analysed by \cite{Holczer2016}. We compared our results with their in the Table~\ref{tab:ttv-old}. Periodic TTV signals of six planets (Kepler-52 d, Kepler-89 c, Kepler-122 e, Kepler-166 c, Kepler-283 c and Kepler-549 c) were already reported by \cite{Thompson2018} but they did not determine any parameters of these changes. 43 of these planets with periodic TTV were not reported in any other paper. Discovering these new TTV systems was a result of using our method of LC de-trending and time of transit measurement. The all planets with unreported periodic TTV and six planets reported (but not analysed) by \cite{Thompson2018} are listed in Table~\ref{tab:ttv} with periods and amplitudes of TTV changes. We used sinusoidal model of these variations to determined their period and amplitude. For finding correct values of sinusoidal model's parameters, we ran simple Levenberg-Marquardt algorithm \citep{Marquardt1963}. Amplitudes of found variations vary between approximately 2 and 70 minutes and periods vary from only 80 to more than 1500 days. These changes could be caused by interaction with another body(ies) (e.g. \cite{Agol2005}), stellar activity (e.g. spots) or other effects in studied systems. 

95 exoplanets shows chaotic or abnormal TTVs (note 3 in Table \ref{tab:ephm}). This type of variations is the most probably caused by stellar activity (e.g. \cite{Oshagh2013}) and/or gravitational interaction with another bodies in the systems. Examples of these systems are shown on Figure~\ref{fig:oc-chaotic}.

\section*{Acknowledgement}
This work was supported by the Slovak Research and Development Agency under the contract No. APVV-15-0458. M.V. would like to thank the project VEGA 2/0031/18. The research of P.G. was supported by the VVGS-PF-2017-724 internal grant of the Faculty of Science, P. J. \v{S}af\'{a}rik University in Ko\v{s}ice.\\

\section*{Supporting information}
Additional Supporting Information may be found in the on-line version of this article:\\
\noindent Table \ref{tab:ephm}. The new linear ephemeris of Kepler exoplanets.


\begin{thebibliography}{}
\bibitem[Agol et al.(2005)]{Agol2005}Agol, E., Steffen, J., Sari, R., \& Clarkson, W.\ 2005, \mnras, 359, 567 
\bibitem[Akeson et al.(2013)]{akeson2013}Akeson, R.~L., Chen, X., Ciardi, D., et al.\ 2013, \pasp, 125, 989
\bibitem[Borucki et al.(2010)]{borucki2010}Borucki, W.~J., Koch, D., Basri, G., et al.\ 2010, \sci, 327, 977
\bibitem[Fabrycky et al.(2012)]{Fabrycky2012} Fabrycky, D.~C., Ford, E.~B., Steffen, J.~H., et al.\ 2012, \apj, 750, 114 
\bibitem[Gajdo\v{s} et al.(2017)]{kep410}Gajdo{\v s}, P., Parimucha, {\v S}., Hamb{\'a}lek, {\v L}., \& Va{\v n}ko, M.\ 2017, \mnras, 469, 2907
\bibitem[Holczer et al.(2016)]{Holczer2016}Holczer, T., Mazeh, T., Nachmani, G., et al.\ 2016, \apjs, 225, 9
\bibitem[Howell et al.(2014)]{Howell2014}Howell, S.~B., Sobeck, C., Haas, M., et al.\ 2014, \pasp, 126, 398
\bibitem[Mandel \& Agol(2002)]{MA}Mandel, K., \& Agol, E.\ 2002, \apjl, 580, L171 
\bibitem[Marquardt(1963)]{Marquardt1963}Marquardt, D. 1963, SIAM J Appl Math, 11, 431
\bibitem[Oshagh et al.(2013)]{Oshagh2013}Oshagh, M., Santos, N.~C., Boisse, I., et al.\ 2013, \aap, 556, A19.
\bibitem[Sing(2010)]{Sing}Sing, D.~K.\ 2010, \aap, 510, A21 
\bibitem[Steffen et al.(2012)]{Steffen2012} Steffen, J.~H., Fabrycky, D.~C., Ford, E.~B., et al.\ 2012, \mnras, 421, 2342 
\bibitem[Thompson et al.(2018)]{Thompson2018}Thompson, S.~E., Coughlin, J.~L., Hoffman, K., et al.\ 2018, \apjs, 235, 38.
\bibitem[Wang \& Ji(2014)]{Wang2014} Wang, S., \& Ji, J.\ 2014, \apj, 795, 85 
\bibitem[Wang \& Ji(2017)]{Wang2017} Wang, S., \& Ji, J.\ 2017, \aj, 154, 236 
\bibitem[Xie(2014)]{Xie2014}Xie, J.-W.\ 2014, \apjs, 210, 25 
\end{thebibliography}
\end{document}